# AI Plays? δ-Rationality Games with Nash Equilibrium as Special Case*

Fang-Fang Tang, Peking University (fftang2012@126.com)
& Yongsheng Xu Georgia State University (yxu3@gsu.edu)

June 19, 2025

**Abstract.** A distortion function, which captures the payoff gap between a player's actual payoff and her true payoff, is introduced and used to analyze games. In our proposed framework, we argue that players' actual payoff functions should be used to explain and predict their behaviors, while their true payoff functions should be used to conduct welfare analysis of the outcomes.

**Keywords**: δ-rationality, rational value function, distortion value function, δ-rationality Nash equilibrium

*We are greatly indebted to the inspiring discussions with Mr. Dahong Liu, a techno-entrepreneur, especially about the AI developments. We would like to devote this work to late Professors John C. Harsanyi and Reinhard Selten for their insightful teachings.

1. The Conceptual Framework

Are there any equilibrium solutions for games with irrational players? We were asked once by the Founder and CEO of an AI start-up. "Yes", we answered him, "it is exactly what we have been working on". The issue that bothers him is the mixed society of humans and AIs[1], which seems coming before long, not only the current LLMs (Large Language Models) but also the LAMs (Large Action Models) including humanoid robotics. Note that the AI engineers view the LAMs as the rational agents, while we humans as limited rational or even irrational players. When the AIs are being developed from generative to reasoning, and then to action-oriented, the behavior of AIs will be more and more "rational" in the view of the engineers - measurable, computational, actionable, and predictable, towards approaching "rationality" in the limit. The evolution speed of humans seems far behind that of the AIs. When the AIs begin to act in their new roles and may have new identities and rights with obligations, the new society becomes a mixture of such super-rational agents and humans who are mostly of limited rationalities. Some AIs will be the agents or assistants of humans, while some AIs may be independent groups. How will such a society be shared, organized, managed and governed? It seems to be a behavioral game with interactions among groups of agents (in a general sense) with various levels of rationalities. Needless to say, we need to define what "rationality" is. Interestingly, the traditional game theory following the von Neumann and Morgenstern framework seems to be more suitable for the super-rational AI agents, rather than the human groups. We will propose a more generalized framework for the human groups mostly with limited rationalities.

---

[1] Rosemary J. Thomas, Jan 17, 2024, "The Rise of Large Action Models, LAMs: How AI Can Understand and Execute Human Intentions?": "LAMs can interact with the world with intelligence, including people, circumstances change adaptation, and other LAMs". https://medium.com/version-1/the-rise-of-large-action-models-lams-how-ai-can-understand-and-execute-human-intentions-f59c8e78bc09

The main feature of our generalized framework is that we redefine the payoff function of a player in an innovative way conceptually. It is fair to say that the conventional payoff function of a player in economics and in game theory is typically interpreted as a representation of the player's "all-things-considered" preferences (Baigent 1995, p. 92) over outcomes. As such, no assumptions are being made regarding the motives/factors underlying a player's ranking or preferences over the outcomes. In economics and in game theory, preferences are asked to play a dual role: to explain observed behaviors and to predict future behaviors on the one hand and to conduct welfare analysis on the other hand. To conduct welfare analysis, a player's welfare is viewed as his/her utility, which is widely interpreted as the player's preference satisfaction or desire fulfillment in modern microeconomic theory. However, as noted earlier, a player's preferences are taken as the player's "all-things-considered" preferences, and the degree of preference satisfaction may not adequately capture the player's utility-based conception of welfare and may lead to misleading and wrong welfare analysis as the player's "preferences may be distorted by factual errors, ignorance, careless thinking, rash judgment, or strong emotions hindering rational choice" (Harsanyi, 1977a, pp. 29-30).[2] Faced with such problems, Harsanyi (1977a, 1977b) has suggested to distinguish between a player's "*explicit* preferences, i.e., his preferences as they actually *are* … and his 'true' preferences, i.e., his preferences as they *would* be under 'ideal conditions'."

The main purpose of this paper is two-fold: first, following the suggestion by Harsanyi (1977a, 1977b), we define a player's payoff, which is viewed as a representation of the player's explicit/actual preferences, as having two components: the utility payoff as representing the player's "true" preferences (to be called "the rational value function" following von Neumann and Morgenstern, 2004 edition) and the "distortion" payoff as reflecting the part of the player's non-true preferences (to be called "the distortion value function"); and secondly, we apply our approach to a player's payoff function to reconsider strategic form games.

More formally, we consider a strategic form game with $n > 1$ players and a player $i$'s finite strategy set being given by $S_i$. For a player $i$, let $i$'s payoff (representing the player's actual preferences), $p_i$, consist of two parts[3]: Part one is the "utility" payoff $U_i$ (representing the player's "true" preferences), while Part two is the "distortion" payoff $D_i$ (reflecting the part of the player's non-true preferences). Like the rational value function $U_i$, the distortion value function $D_i$ is also a mapping of all *n*-tuples of pure or mixed strategies into the real numbers and represents what the player will bring into the game in a sense of "irrationality", rather than any "true-preference" satisfaction. For example, in many parts of the world, there are tragic events that some individuals go to streets to hurt or even kill people randomly whom he/she did not know at all (such as driving a car or a truck). This sort of behavior is totally irrational from any perspective of traditional "utility". Rather, such act of destruction seems to be a leak of "anger" for destruction *per se*. We shall assume the "value" of such irrationality can also be expressed by real numbers, in line with the utility measurement, for simplicity.

---

[2] Max Weber (1922; 1978, p..6) had similar observations as he noted that emotions like "anxiety, anger, ambition, envy, jealousy, love, enthusiasm, pride, vengefulness, loyalty, devotion" might influence one's rational choice.

[3] In the vast literature of experimental and behavioral economics, there has been attempts to investigate various factors/concerns/motives such as fairness (Rabin, 1993), inequality aversion (Fehr and Schmidt, 1999), etc. lumped into a player's preferences; see, for example, Fehr and Schmidt (2006) for various studies of such factors/concerns/motives.

The idea of "irrationality" mentioned above is not about intransitive preferences of a player; rather, it refers to some particular behaviors or concerns or values that are deemed to distort the player's "true" preferences [4] and/or sometimes to be destructive to both the player himself/herself and others—behaviors that harm others without benefiting oneself.

Our approach may thus be viewed as making a suggestion for a research paradigm shift: we should use the payoff function $p_i$ to understand/explain and predict the behaviors of a player $i$, but when it comes to analyze $i$'s welfare, we should use $i$'s rational value function/utility payoff $U_i$. Thus, the alleged "burden" of the dual role of a player $i$'s "all-things-considered" preferences can be divided into and shared between $p_i$ and $U_i$: to explain and understand the behavior, we should use $i$'s payoff function $p_i$, and, to analyze $i$'s welfare, we should use $i$'s rational value function $U_i$.

Thus, the payoff function $p_i$ of a player $i$ is given by $p_i(s_1, \ldots, s_n) = \delta_i U_i(s_1, \ldots, s_n) + (1 - \delta_i) D_i(s_1, \ldots, s_n)$, with $s_i$ as a strategy (pure or mixed) of player $i$ and where $\delta_i \in [0,1]$ with $\delta_i$ being interpreted as player $i$'s "degree of rationality". When $\delta_i = 1$ for all $i$, we are back to the traditional game with all the rational analysis of behaviors and conventional welfare analysis. When $\delta_i = 0$ for all $i$, this game will go to the other extreme of "full irrationality" such as everybody going out to hurt unknown people randomly. In between, when $0 < \delta_i < 1$ for some $i$, we have a game of "limited rationality", more or less. Therefore, one may call this index $\delta$ as an index of rationality, in a general sense. Thus, we call our approach here $\delta$-rationality.

Note that $\delta$-rationality is different from "the bounded rationality" proposed by Simon (1957), which was mainly characterized by "search for alternatives, satisficing, and aspiration adaptation" as Selten (1999, p.14) summarized. Our $\delta$-rationality, on the other hand, is more aligned with Selten's (1999, p.35) description "that a person is subdivided into several components that may be in conflict with each other", while we see no point why players do not maximize the available payoffs for themselves. Our approach has the advantage that the formal modelling process is manageable and all the conventional game-theoretical analysis remains valid as a special case of our framework, that is, the vast game-theoretical toolbox developed in the past decades remains to be our effective aide rather than being thrown away completely.

For simplicity, we assume that all the parameters are set before the game starts and this information is publicly known among the players. Therefore, this is a game of complete information and not a game of incomplete information in the sense of Harsanyi (1967-8).

---

[4] This idea of irrationality is similar to irrational behaviors of a person that Max Weber (1922; 1978) had in mind when he commented on how individuals may behave by irrational factors such as those mentioned in the footnote 2 above.

**Theorem:** Every generalized finite game has a Nash equilibrium point for each *n*-tuple of $\delta_i \in [0,1]$, $i=1, ..., n$, to be called the δ-rationality Nash equilibrium.

Proof: Analogous to Nash (1950, 1951), to be outlined briefly as follows.

For each $\delta_i \in [0,1]$, $i=1, ..., n$, any *n*-tuple of strategies (one for each player) is a point in the product space obtained by multiplying the *n* strategy spaces of the *n* players. One such *n*-tuple counters another if the strategy of each player in the countering *n*-tuple yields the highest obtainable expectation for its player against the *n*-1 strategies of the other players in the countered *n*-tuple. Obviously, a self-countering *n*-tuple is an equilibrium point.

The correspondence of each *n*-tuple with its set of countering *n*-tuples gives a one-to-many mapping of the product space into itself. From the definition of countering, one can see that the set of countering points of a point is convex. Since the payoff functions are continuous, the graph of the mapping is closed. From the closedness of the graph and the image of each point under the mapping being convex, the mapping has a fixed point by Kakutani (1941) theorem, that is, an equilibrium point.

**Remark 1.** There is a possible continuum of Nash equilibrium points for every generalized finite game for $\delta_i \in [0,1]$, $i=1, ..., n$.

**Remark 2.** If $\delta_i = \bar{\delta}$ (a constant) for all *i*, the generalized game has Nash equilibrium points all of which as a function of the constant rationality index $\bar{\delta}$. If all players have the same rationality index $\bar{\delta}$, we shall call such a generalized game as a constant-rationality game. The extreme cases are $\bar{\delta}=1$, back to the traditional game type, and $\bar{\delta}=0$, the fully irrational game.

**Remark 3.** The distribution characteristics of $\delta_i$, when $\delta_i$ is viewed as a random variable on [0, 1], $i=1, ..., n$, will set distribution characteristics of equilibrium points. We shall call such a generalized game as a variable-rationality game. For example, if $\delta_i \in [0,1]$ follows a normal distribution for all *i*, we may call such a generalized game as a game of normal distribution rationality.

Note that our generalized game is different from a stochastic game where the play proceeds by steps from position to position, according to transition probabilities controlled jointly by the players (see Shapley 1953).

## 2. Numerical Example I

We shall give some numerical examples for illustrative purpose as follows.

A quote first, from the iPhone designer, Fadell (2022) part VI:
"I can't explain it. Just as I never learned their true reasons for selling Nest, I never heard an explanation for why they decided to keep it. Maybe the fact that Amazon was interested made

Larry realize that Nest was a valuable asset after all. Maybe it was all an elaborate *game of chicken* to get me to toe the line and cut costs. Maybe they never had a real plan to begin with and this all happened because of some exec's casual whim. You'd be surprised how often that's the reason behind major changes.

People have this vision of what it's like to be an executive or CEO or leader of a huge business unit. They assume everyone at that level has enough experience and savvy to at least appear to know what they're doing. They assume there's thoughtfulness and strategy and long-term thinking and reasonable deals sealed with firm handshakes.

But some days, it's high school. Some days, it's kindergarten." (italics added)

First, the traditional utility payoff matrix of the chicken game:

|  |  | Player 2 | |
|---|---|---|---|
|  |  | swerve | Straight |
| Player 1 | swerve | 1, 1 | -2, 2 |
|  | straight | 2, -2 | -4, -4 |

which has two pure strategy Nash equilibria (swerve, straight) and (straight, swerve), and one mixed strategy Nash equilibrium $\{(\frac{2}{3}, \frac{1}{3}), (\frac{2}{3}, \frac{1}{3})\}$.

If the distortion value payoff matrix of this game is:

|  |  | Player 2 | |
|---|---|---|---|
|  |  | swerve | Straight |
| Player 1 | swerve | 0, 0 | 0, x>0 |
|  | straight | 0, 0 | 0, x>0 |

That is, player 1 has zero distortion value for either "swerve" or "straight", while player 2 has zero distortion value for "swerve" but a positive distortion value of x>0 for "straight" (a leak of "anger" with, for instance, any suicidal behavior).

For $\delta_i \in [0,1]$ (with i=1,2), the δ-rationality game will be:

|  |  | Player 2 | |
|---|---|---|---|
|  |  | swerve | straight |
| Player 1 | swerve | $\delta_1, \delta_2$ | $-2\delta_1, 2\delta_2+x(1-\delta_2)$ |
|  | straight | $2\delta_1, -2\delta_2$ | $-4\delta_1, -4\delta_2+x(1-\delta_2)$ |

Because $\delta_2 < 2\delta_2 + x(1-\delta_2)$ for any $\delta_2 \in [0,1]$ with x>0, "straight" will be a dominant strategy for player 2 as long as $-2\delta_2 < -4\delta_2 + x(1-\delta_2)$, i.e., $\delta_2 < \frac{x}{2+x}$. Therefore, there is only one pure strategy

Nash equilibrium (swerve, straight) for this δ-rationality game when $\delta_2 < \frac{x}{2+x}$ and $\delta_1 > 0$.

From the above discussion, it is clear that "straight" is a dominant strategy for player 2 as long as $\delta_2 < \frac{x}{2+x}$. $\frac{x}{2+x}$ can be viewed as the upper bound of $\delta_2$ such that, for any given distortion value payoff $x$, if player 2's degree of rationality is less than this upper bound ($\delta_2 < \frac{x}{2+x}$), then "straight" is a dominant strategy for player 2. Noting that $\lim_{x \to +\infty} (\frac{x}{2+x}) = 1$, we can have the scenario in which an "almost fully rational" ($\delta_2$ is close to 1) player 2 may find that "straight" is a dominant strategy when the distortion value payoff $x$ becomes very large. Furthermore, since $\frac{d}{dx}(\frac{x}{2+x}) = \frac{2}{(2+x)^2} > 0$, it suggests that the upper bound of $\delta_2$ strictly increases with the increase of x, as follows for instance:

| | |
|---|---|
| $x = \frac{1}{2}$ | $\delta_2 < \frac{1}{5}$ |
| $x = 1$ | $\delta_2 < \frac{1}{3}$ |
| $x = \frac{3}{2}$ | $\delta_2 < \frac{3}{7}$ |
| $x = 2$ | $\delta_2 < \frac{1}{2}$ |
| $x = 3$ | $\delta_2 < \frac{3}{5}$ |
| $x = 4$ | $\delta_2 < \frac{2}{3}$ |

3. Numerical Example II

For the well-known "Prisoner's Dilemma" game, we take the simple utility payoff matrix in Maschler et al. (2013), p.259, with D=Defect and C=Cooperate,

Player 2

| | | D | C |
|---|---|---|---|
| Player 1 | D | 1, 1 | 4, 0 |
| | C | 0, 4 | 3, 3 |

which has only one Nash equilibrium (D, D) in dominant pure strategies.

If the distortion value payoff matrix of this game is

|  | | Player 2 | |
|---|---|---|---|
|  | | D | C |
| Player 1 | D | -1, -1 | -1, 0 |
|  | C | 0, -1 | 0, 0 |

That is, for these two players, both of them have zero distortion value for C, but a "-1" distortion value for D, since they happen to be close friends who are loyal to each other and would feel very guilty to betray their friend. One can interpret this phenomenon as thieves may have friends and loyalty as well, if not more than other people[5]. There are numerous betrayal events in human history, but we have also witnessed many loyalty and friendship cases as well. Human nature has its bright side and dark side, just as a coin with two sides.

For $\delta_1 \in [0,1]$ and $\delta_2 \in [0,1]$, the δ-rationality game will be

|  | | Player 2 | |
|---|---|---|---|
|  | | D | C |
| Player 1 | D | $2\delta_1-1, 2\delta_2-1$ | $5\delta_1-1, 0$ |
|  | C | $0, 5\delta_2-1$ | $3\delta_1, 3\delta_2$ |

One can see easily that as long as $2\delta_1-1<0$ and $5\delta_1-1<3\delta_1$, i.e. $\delta_1<\frac{1}{2}$, "C" is a strictly dominant strategy for player 1. Similarly, when $2\delta_2-1<0$ and $5\delta_2-1<3\delta_2$, i.e. $\delta_2<\frac{1}{2}$, "C" is a strictly dominant strategy for player 2. Clearly, the unique pure strategy equilibrium will be (C, C) for either $\delta_1<\frac{1}{2}$ and $\delta_2<\frac{1}{2}$. That is, so long as both of them are less than half-rational, they will end up in the best solution[6] (C, C) for this generalized δ-rationality game.

On the other hand, when $2\delta_1-1>0$ and $5\delta_1-1>3\delta_1$, i.e. $\delta_1>\frac{1}{2}$, "D" is a strictly dominant strategy for player 1. Similarly, when $2\delta_2-1>0$ and $5\delta_2-1>3\delta_2$, i.e. $\delta_2>\frac{1}{2}$, "D" is a strictly dominant strategy for player 2. The unique Nash equilibrium will be (D, D) when $\delta_1>\frac{1}{2}$ and $\delta_2>\frac{1}{2}$. More than half-rational or what people usually say "you think too much" (particularly the intelligent ones) seems making no good for the players of this generalized δ-rationality game.

---

[5] There is a proverb saying that "There is honor among thieves.", which can be traced back to the writings by Cicero (De Officiis) 2000 years ago in the early Roman Empire.

[6] Here, the best solution is defined in terms of their "true" preferences as represented by their rational value functions.

It is easy to check that, when $\delta_1 > \frac{1}{2}$ and $\delta_2 < \frac{1}{2}$, the unique Nash equilibrium will be (D, C); when $\delta_1 < \frac{1}{2}$ and $\delta_2 > \frac{1}{2}$, the unique Nash equilibrium will be (C, D). That is, player 1 would have a much better payoff (i.e., $5\delta_1-1$) than player 2 (i.e., 0) if player 1 is more than half-rational but player 2 is less than half-rational. The situation is reversed if player 1 is less than half-rational but player 2 is more than half-rational, with player 2 having a much better payoff (i.e., $5\delta_2-1$) than player 1 (i.e., 0)

We can summarize the Nash equilibrium points and the equilibrium payoffs of the above cases as follows,

Player 2

|  | $\bar{\delta}_2 > \frac{1}{2}$ | $\bar{\delta}_2 < \frac{1}{2}$ |
|---|---|---|
| $\bar{\delta}_1 > \frac{1}{2}$ | (D, D): $(2\delta_1-1, 2\delta_2-1)$ | (D, C): $(5\delta_1-1, 0)$ |
| $\bar{\delta}_1 < \frac{1}{2}$ | (C, D): $(0, 5\delta_2-1)$ | (C, C): $(3\delta_1, 3\delta_2)$ |

Player 1

It can be checked that, when $\delta_1 > \frac{1}{2}$ and $\delta_2 = \frac{1}{2}$, this δ-rationality game will have a continuum of Nash equilibria: Player 1 always plays the pure strategy D and player 2 plays the strategy D with a probability $\pi \in [0, 1]$ and the strategy C with the probability $(1 - \pi)$. Player 1's expected payoff is $[(5 - 3\pi)\delta_1 - 1]$ while player 2's expected payoff is 0. On the other hand, when $\delta_1 < \frac{1}{2}$ and $\delta_2 = \frac{1}{2}$, this δ-rationality game will have another continuum of Nash equilibria: Player 1 always plays the pure strategy C and player 2 plays the strategy D with a probability $\pi \in [0, 1]$ and the strategy C with the probability $(1 - \pi)$. Player 1's expected payoff is $[(1 - \pi)3\delta_1]$ while player 2's expected payoff is 1.5.

Analogously, when $\delta_1 = \frac{1}{2}$ and $\delta_2 > \frac{1}{2}$, this δ-rationality game will have a continuum of Nash equilibria: Player 2 always plays the pure strategy D and player 1 plays the strategy D with a probability $\pi \in [0, 1]$ and the strategy C with the probability $(1 - \pi)$. Player 2's expected payoff is $[(5 - 3\pi)\delta_2 - 1]$ while player 1's expected payoff is 0. But, when $\delta_1 = \frac{1}{2}$ and $\delta_2 < \frac{1}{2}$, this δ-rationality game will have another continuum of Nash equilibria: Player 2 always plays the pure strategy C and player 2 plays the strategy D with a probability $\pi \in [0, 1]$ and the strategy C with the probability $(1 - \pi)$. Player 2's expected payoff is $[(1 - \pi)3\delta_2]$ while player 1's expected payoff is 1.5.

When $\delta_1=\frac{1}{2}$ and $\delta_2=\frac{1}{2}$, or to say, both are half-rational, the δ-rationality game will be

Player 2

|  | D | C |
|---|---|---|
| D | 0, 0 | $\frac{3}{2}$, 0 |
| C | 0, $\frac{3}{2}$ | $\frac{3}{2}$, $\frac{3}{2}$ |

Player 1

which turns out to be a particularly interesting game: Any pure strategy combination will be a pure strategy equilibrium, and any probability combination of the pure strategies will be a mixed strategy equilibrium.

In fact, for any real numbers *a* and *b*, the following game will exhibit the same interesting property that any pure strategy combination will be a pure strategy equilibrium, and any probability combination of the pure strategies will be a mixed strategy equilibrium. Since each player can do whatever he/she likes but the Nash equilibrium will obtain any way, we may call a class of such games as a class of "Easy Games". That is, just go easy and take it easy, do whatever you like. This by-product finding has brought us quite some fun.

Player 2

|  | A | B |
|---|---|---|
| A | *a, a* | *b, a* |
| B | *a, b* | *b, b* |

Player 1

There is another special feature which needs particular attention: The outcome of this Easy Game depends on the choice of the other player, no matter what choice the player makes. For instance, if Player 1 chooses pure strategy A, the outcome of this game completely depends on whether Player 2 chooses A or B, since either the outcome (A, A) or (A, B) is a Nash equilibrium point. In some sense, the "fate" (outcome) of Player 1 is totally up to what Player 2 decides, i.e., the decision of Player 2 dictates the "fate" of Player 1. This is inherently different from the ordinary games where the interactive decisions of all players determine the outcome.

## 4. AI Plays?

The fast progress in AI, from LLMs to LAMs, is giving us a new perspective in playing games.

For the LLMs, as Stephen Wolfram points out:
"At each step it gets a list of words with probabilities. But which one should it actually pick to add to the essay (or whatever) that it's writing? One might think it should be the "highest-ranked" word (i.e. the one to which the highest "probability" was assigned). But this is where a bit of voodoo begins to creep in. Because for some reason—that maybe one day we'll have a scientific-style understanding of—if we always pick the highest-ranked word, we'll typically get a very "flat" essay, that never seems to "show any creativity" (and even sometimes repeats word for word). But if sometimes (at random) we pick lower-ranked words, we get a "more interesting" essay. The fact that there's randomness here means that if we use the same prompt multiple times, we're likely to get different essays each time. And, in keeping with the idea of voodoo, there's a particular so-called "temperature" parameter that determines how often lower-ranked words will be used, and for essay generation, it turns out that a "temperature" of 0.8 seems best. (It's worth emphasizing that there's no "theory" being used here; it's just a matter of what's been found to work in practice. And for example the concept of "temperature" is there because exponential distributions familiar from statistical physics happen to be being used, but there's no "physical" connection—at least so far as we know.)."[7]

We humans have a capability, more or less, to "sense" or "feel" how rational or irrational the other players may be. For instance, one person's feeling about the reliability of another person may be captured by the underlying meaning of various expressions such as "nuts" or "this guy is crazy" that we often use in daily life. On the other hand, human brains may not be powerful enough in collecting and processing all possible information and may well be biased by various emotional sentiments to cause disastrous consequences (a point also observed by Weber (1922; 1978) as we cited in the footnotes 2 and 4 earlier). AI systems, either as an observer or as a player, seem to evolve into powerful tools to master the massive parameters in games which need to be processed dynamically and instantly. Human players have a "feeling" about any game roughly to form subjective beliefs, such as how much "craziness" there may be and what to get therewith from, for instance, the AI analysis on the social media information. This effort is to sense the players' attributes such as intentions, sentiments and so on, not just to understand the relevant expressions, that is, more in "instinct" than by languages.

One extremely important issue in the LAMs (Execute AI or Action-oriented AI) is intent-recognition, to identify the intentions of the users or players, which is also a crucial element in playing games, not only for commercial use. The current LLMs are only at the generative and reasoning stages yet, for conversations but not for actions. For conversations, LLMs have temperatures around 0.8 best for generating next words, not picking the word with the highest probability. But the Action-oriented AI must have accurate intent-recognition and choose the optimal option available at each step, in this sense, a fully rational agent or to say the fully rational economic "man", while we humans are mostly of limited rationality on the contrary[8].

Our approach suggests a research paradigm shift that it is no longer necessary to assume full rationality in the fields of game theory, economics and other disciplines. In real life, various kinds of humans have so many types and degrees of "rationality" which need to be indexed rather than ignored. That is, textbooks should be modified if their assumptions are inappropriate, instead of selecting the tiny "rational" part of real life for modelling.

---

[7] Stephen Wolfram (2023), What Is ChatGPT Doing … and Why Does It Work?
https://writings.stephenwolfram.com/2023/02/what-is-chatgpt-doing-and-why-does-it-work/

[8] "If you're making a self-driving car, you don't want to assume that all the other drivers on the road are perfectly rational, and going to behave optimally," Dr. Noam Brown at Meta AI, DeepMind's Latest AI Trounces Human Players at the Game 'Stratego', https://singularityhub.com/2022/12/05/deepminds-latest-ai-trounces-human-players-at-the-game-stratego/ . We are grateful to Dahong Liu for suggesting the above quote of Dr. Brown.

One may easily understand that players for any game can be of various rationality levels or, to say, various degrees of $\delta_i$, and our framework of generalized games can be used to model this general scenario. Simple games can be solved "manually" by humans, but any large complex game will need AI systems to "learn" the parameters and "select" the solutions, particularly for dynamic and instantly changing situations. Equilibrium selection is an extremely difficult task even for the traditional game theory with full rationality assumption (see Harsanyi and Selten 1988). However, in real life, any game will always produce an outcome, whatever result may be. By our framework, AI systems can be used to predict the outcome or the most likely result through indexing and ranking, no matter what rationality levels of the players are.

It is not unimaginable that games may be played among human players, or among humans and AI systems (including AI-driven robots), or even among AI systems, in the future. From this point of view, a more general framework of game theory will be urgently needed, indeed. Certainly, another conceptual difficulty will emerge: What is the payoff function of AIs? We will have to leave this difficult conceptual issue to future philosophical discussion, but explore other equilibrium concepts and the tough issue of incomplete information (for human players) next with much added mathematical complexity in separate papers.